\documentclass[twoside]{dis08}
\usepackage[latin1]{inputenc}
\usepackage[dvips]{graphicx,epsfig,color}
\usepackage{wrapfig,rotating}
\usepackage{amssymb,amsmath,array}

\def\lsim{\raise0.3ex\hbox{$<$\kern-0.75em\raise-1.1ex\hbox{$\sim$}}}
\def\gsim{\raise0.3ex\hbox{$>$\kern-0.75em\raise-1.1ex\hbox{$\sim$}}}

\begin{document}
\def\be{\begin{equation}}
\def\ee{\end{equation}}

\title{The Ratio of {\boldmath$\sigma_L(W^2, Q^2)/\sigma_T(W^2, Q^2)$}
 in the Color
Dipole Picture\footnote{Talk presented at DIS 2008, London}}

\author{Dieter Schildknecht
%
\thanks{Supported by Deutsche Forschungsgemeinschaft, contract number
SCHI 189/6-2}
%
\vspace{.3cm}\\
%
Fakult{\"a}t f{\"u}r Physik - Universit{\"a}t Bielefeld \\
Universit{\"a}tsstrasse 25 - D-33615 Bielefeld\\ and\\
Max-Planck Institute f{\"u}r Physik (Werner-Heisenberg-Institut) \\
F{\"o}hringer Ring 6 - D-80805 M{\"u}nchen 
%
}

\maketitle

\begin{abstract}
The transverse size of $q \bar q$ fluctuations of the longitudinal
photon is reduced relative to the transverse size of $q \bar q$
fluctuations of the transverse photon. This implies $R (W^2, Q^2) =
0.375$ or, equivalently, $F_L(W^2,Q^2)/F_2(W^2,Q^2) = 0.27$ 
for $x \cong Q^2/W^2 \ll 1$ and $Q^2$ sufficiently large, 
while $R (W^2, Q^2) = 0.5$, if
this effect is not taken into account. Forthcoming experimental data from
HERA will allow to test this prediction.
\end{abstract}

In this written version of my talk,
I will restrict myself to a brief summary of our prediction on the ratio
of the longitudinal to the transverse photoabsorption cross section. I
refer to the recent paper \cite{Ku-Schi} for details and a more complete list
of references.

At low values of $x \cong Q^2/W^2 << 1$, in terms of the imaginary part
of the virtual Compton-scattering amplitude, deep inelastic scattering
(DIS) proceeds via forward scattering of (timelike) quark-antiquark,
$q \bar q$, fluctuations of the virtual spacelike photon on the proton.
In its interaction with the proton, a $q \bar q$ fluctuation acts as 
a color dipole. A massive $q \bar q$ fluctuation is identical to the
$(q \bar q)^{J=1}$ vector state originating from a timelike photon in
$e^+e^-$ annihilation at an $e^+e^-$ energy equal to the mass, $M_{q \bar q}$,
of the $q \bar q$ state.

Validity of the color-dipole picture (CDP) requires the lifetime of a 
$q \bar q$ fluctuation, $L$, to be large,
\be
L = \frac{W^2}{x + \frac{M^2_{q \bar q}}{W^2}} \cdot \frac{1}{M_p}
\equiv L_0 \frac{1}{M_p},
\label{1}
\ee
i.e.
\be
L_0 = \frac{1}{x + \frac{M^2_{q \bar q}}{W^2}} \gg 1.
\label{2}
\ee
Besides $x \ll 1$, at any given $\gamma^*p$ center-of-mass energy, $W$,
the mass of the contributing $q \bar q$ states, $M_{q \bar q}$, is
limited. In view of the subsequent discussions, we note the relation 
between the $q \bar q$ mass and the transverse momentum of the
quark (antiquark), $\vec k_\bot$, that is given by
\be
M^2_{q \bar q} = \frac{\vec k^{~2}_\bot}{z(1-z)},
\label{3}
\ee
where $0 \le z \le 1$ denotes the usually employed variable that is
related to the $q \bar q$ rest-frame angle between the $\gamma^*p$ axis
and the three-momentum of the (massless) quark,
\be
\sin^2 \vartheta = 4 z (1-z).
\label{4}
\ee

The coupling strength of a timelike photon of mass $M_{q \bar q}$ to
a $q \bar q$ state of mass $M_{q \bar q}$ is determined by the longitudinal
and transverse components of the electromagnetic current \cite{Cvetic},
\be
\sum_{\lambda = - \lambda^\prime = \pm 1} \vert j_L^{\lambda, \lambda^\prime}
\vert^2 = 8 M^2_{q \bar q} z (1-z)
\label{5}
\ee
and
\be
\sum_{\lambda = - \lambda^\prime = \pm 1} \vert j_T^{\lambda, \lambda^\prime}
(+) \vert^2 = \sum_{\lambda = - \lambda^\prime = \pm 1} 
j_T^{\lambda, \lambda^\prime} (-) \vert^2 = 2 M^2_{q \bar q} (1 - 2 z (1-z)).
\label{6}
\ee
where $j_T^{\lambda, \lambda^\prime} (+)$ and 
$j_T^{\lambda, \lambda^\prime} (-)$ refer to positive and negative helicity
of the transverse photon. The $q \bar q$ pair consists of a quark and antiquark
of opposite helicity.

From (\ref{5}) and (\ref{6}), we see that the transition of a longitudinal
photon to a $q \bar q$ pair prefers $z \not= 0$, while a transverse photon
prefers $z = 0$. Transverse photons produce (relatively) small-$k_\bot$
pairs of mass $M_{q \bar q}$ according to (\ref{3}), while longitudinal 
photons produce (relatively)
large-$k_\bot$ pairs. From (\ref{3}) with (\ref{5}) and (\ref{6}), one
finds that the ratio of the average transverse momenta is given 
by \cite{Ku-Schi}
\be
\rho = \frac{\langle \vec k^{~2}_\bot\rangle_L}{\langle \vec k^{~2}_\bot \rangle_T}
= \frac{4}{3}.
\label{7}
\ee
From the uncertainty relation, the ratio of the effective transverse sizes
is then given by
\be
\frac{\langle \vec r^{~2}_\bot\rangle_L}{\langle \vec r^{~2}_\bot \rangle_T}
= \frac{1}{\rho} = \frac{3}{4}.
\label{8}
\ee
Longitudinal photons, $\gamma^*_L$, produce ``small-size'' pairs, while
transverse photons, $\gamma^*_T$, produce ``large-size'' pairs. The ratio of
the average sizes is given by (\ref{8}).\footnote{In my presentation at
DIS 2008 I incorrectly stressed helicity independence, $\rho = 1$, as
a necessity. See, however, ref. \cite{DIS07}, where helicity independence
was introduced as a hypothesis.}

The transition from a timelike photon interacting with the proton via a
$q \bar q$ pair of mass $M_{q \bar q}$ to a spacelike photon fluctuating into 
a mass continuum of $q \bar q$ vector states is provided by the 
CDP \cite{Nikolaev}. In a
formulation that expresses the photoabsorption cross section  in terms
of the scattering of $(q \bar q)^{J=1}_{L,T}$ longitudinal and transverse 
$q \bar q$ vector states, one obtains
\be
\sigma_{\gamma^*_{L,T}} (W^2, Q^2) = \frac{2 \alpha R_{e^+e^-}}{3 \pi^2}
Q^2 \int d^2 r^\prime_\bot K^2_{0,1} (r_\bot^\prime Q) 
\sigma_{(q \bar q)^{J=1}_{L,T} p}
(r^\prime_\bot, W^2).
\label{9}
\ee
where, taking care of the size effect (\ref{8}),
\be
    \sigma_{(q \bar q)^{J=1}_T p} (r^\prime_\bot, W^2) = \rho 
    \sigma_{(q \bar q)^{J=1}_L p} (r^\prime_\bot, W). 
   \label{10}
\ee
In (\ref{9}) and (\ref{10}), $r^\prime_\bot$ is related to the
transverse size of the $q \bar q$ pair by
\be
\vec r^{~\prime}_\bot = \vec r_\bot \sqrt{z(1-z)}.
\label{11}
\ee
To incorporate the coupling of the $q \bar q$ pair to two gluons,
the $(q \bar q)^{J=1}$ interaction in (\ref{10}) has to vanish as
$\vec r^{~\prime 2}_\bot$ for 
$\vec r^{~\prime 2}_\bot \to 0$ (``color transparency''). Due
to the strong decrease of the modified Bessel functions $K_{0,1} 
(r_\bot^\prime Q)$ for large values of $r^\prime_\bot Q$, the integral 
in (\ref{9}) for large $Q^2$ is dominated by $r^{\prime 2}_\bot \to 0$,
and, accordingly, from (\ref{9}) with (\ref{10}) and color transparency,
we have for $x \ll 1$ and $Q^2$ sufficiently large,
\be
R(W^2, Q^2) \equiv \frac{\sigma_{\gamma^*_L p}(W^2, Q^2)}{\sigma_{\gamma^*_T p}
 (W^2, Q^2)} =
\frac{\int d^2 r^\prime_\bot r^{\prime 2}_\bot K^2_0 (r^\prime_\bot Q)}
{\rho \int d^2 r^\prime_\bot r^{\prime 2}_\bot K^2_1 (r^\prime_\bot Q)} = 
\frac{1}{2\rho} = \frac{3}{8} = 0.375.
\label{12}
\ee
Equivalently, in terms of the structure functions,
\be
     \frac{F_L (W^2, Q^2)}{F_2 (W^2, Q^2)} = \frac{1}{1+2 \rho} = 
     \frac{3}{11} \simeq 0.27.
\label{13}
\ee
In (\ref{12}) and (\ref{13}), the equality \cite{Gradsteyn}
\be
\int^\infty_0 dy~y^3 K^2_0 (y) = \frac{1}{2} \int^\infty_0 dy~y^3 K^2_1 (y)
\label{14}
\ee
was used, and the value of $\rho = \frac{4}{3}$ from (\ref{8}) was
inserted. The predictions (\ref{12}) and (\ref{13}) are independent of a
specific ansatz for the color-dipole cross section. They rely on the CDP
in the $r^\prime_\bot$ representation (\ref{9})
combined with color transparency and the
$q \bar q$ transverse-size effect incorporated into the proportionality
(\ref{10}).

The parameter $\rho$ from (\ref{10}), making use of the first equality in
(\ref{13}), can be 
determined from measurements of DIS at different electron-proton center-of-mass
energies, $\sqrt s$, for fixed values of $x$ and $Q^2$. The reduced cross
section of DIS is given by
\be
\sigma_r (x, y, Q^2) = F_2 (x, Q^2) 
\left(1 - \frac{y^2}{1 + (1-y)^2} \frac{1}{1 + 2 \rho} \right),
\label{15}
\ee
where $y = Q^2/xs$. The slope of a straight-line fit of $\sigma_r
(x,y,Q^2)$ as a function of $0 \le y^2/(1+(1-y)^2)\le 1$ determines
$\rho$. A value of
\be
\rho = 1
\label{16}
\ee
corresponds to helicity independence, i.e. equality of the forward-scattering
amplitudes of $(q \bar q)^{J=1}_h$ fluctuations of the photon on the
proton for helicities $h = 0, h = + 1$ and $h = -1$. A deviation from
$\rho = 1$ rules out helicity independence. The reduced transverse size
of longitudinally polarized $(q \bar q)^{J=1}$ states relative to 
transversely polarized $(q \bar q)^{J=1}$ states implies a value of
$\rho = 4/3$.

The preliminary results from HERA on the measurements of $F_L (W^2, Q^2)$ 
presented \cite{DIS2008}
at DIS 2008 seem to disfavor helicity independence. The measurements were
carried out at values of $Q^2$ and $W^2$ at which $F_2(W^2, Q^2) \cong
1.2$.
According to (\ref{13}),
one finds $F_L \cong 0.4$ for $\rho = 1$ and $F_L \cong 0.33$ for
$\rho = 4/3$. The prediction of $F_L \cong 0.33$ seems consistent 
with the data presented
at DIS 2008. 

An interesting upper bound on $R(W^2, Q^2)$ was recently derived 
\cite{Ewerz} in the usual formulation of the CDP. The bound is
given by 
\be
R(W^2, Q^2) \le 0.37248,
\label{17}
\ee
or, in terms of $\rho$,
\be
\rho \gsim 1.34.
\label{18}
\ee
The bound is inconsistent with helicity independence, $\rho = 1$, and,
strictly speaking, with our prediction of $\rho = 1.33$ from (\ref{7}).

The usual CDP that implies the bound (\ref{17}) is not explicitly 
formulated in terms of $(q \bar q)^{J=1}$
vector state scattering and, in particular, it contains an 
$\vec r_\bot$-dependent dipole cross section that is independent of
$z(1-z)$. If a $z(1-z)$ dependence is allowed, the derivation of the
upper bound (\ref{17}) fails.

\begin{wrapfigure}{r}{0.5\columnwidth}
\centerline{\includegraphics[width=0.45\columnwidth]{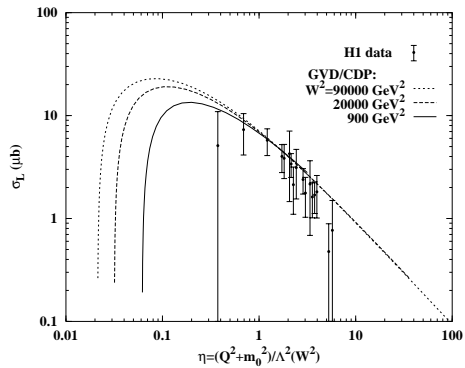}}
\caption{The predictions from the ansatz for the color-dipole cross
section in ref. \cite{Surrow} are compared \cite{Tentyukov} 
with H1 data extracted from measurements of $F_2(W^2, Q^2)$ under
theoretical input assumption.}\label{Fig:1}
\end{wrapfigure}

An example for a $z(1-z)$-dependent dipole cross section is given by 
our ansatz \cite{Surrow}. It contains helicity independence  and provides a
successful representation of the experimental data for the total
photoabsorption cross section. In fig. 1, we show the longitudinal 
photoabsorption cross section \cite{Tentyukov} compared with data available
at the time \cite{H1}. Under theoretical input assumptions, the data were
extracted by the H1 collaboration from the
measured structure function $F_2 (W^2, Q^2)$. There is a tendency for
the theoretical prediction, based on $\rho = 1$, to overestimate the data,
thus requiring $\rho > 1$.

The starting point of the present work is the CDP in a representation 
(``$r^\prime_\bot$-representation'') that explicitly factorizes $\gamma^*p$
scattering into three distinct steps, $\gamma^* (q \bar q)^{J=1}_{L,T}$
coupling, propagation and scattering on the proton. Independently of the
specific prediction on the parameter $\rho$, it is worth to be stressed
that the separation data directly, via measurement of $\rho$, determine
the relative magnitude of the scattering of longitudinally versus
transversely polarized massive $(q \bar q)^{J=1}$ vector states on the proton.

\begin{footnotesize}

\end{footnotesize}


\end{document}